# A Novel Error Correcting System Based on Product Codes for Future Magnetic Recording Channels

Vo Tam Van and Seiichi Mita, *Members, IEEE*

Toyota Technological Institute, Hisakata, Tempaku, Nagoya 468-8511, Japan

**We propose a novel construction of product codes for high-density magnetic recording based on binary low-density parity check (LDPC) codes and binary image of Reed Solomon (RS) codes. Moreover, two novel algorithms are proposed to decode the codes in the presence of both AWGN errors and scattered hard errors (SHEs). Simulation results show that at a bit error rate (bER) of approximately $10^{-8}$, our method allows improving the error performance by approximately 1.9dB compared with that of a hard decision decoder of RS codes of the same length and code rate. For the mixed error channel including random noises and SHEs, the signal-to-noise ratio (SNR) is set at 5dB and 150 to 400 SHEs are randomly generated. The bit error performance of the proposed product code shows a significant improvement over that of equivalent random LDPC codes or serial concatenation of LDPC and RS codes.**

## I. INTRODUCTION

For future magnetic recording systems, bit patterned media (BPM) and shingled writing recording (SWR) are two promising candidates in order to implement high-density recording of more than 10 Tera-bits per square inch. From the view point of error correction, these systems will include not only conventional errors such as random errors due to the additive white Gaussian noise (AWGN) and burst errors due to media defects, but scattered hard errors (SHEs) as well. SHEs are errors with large changes in signal amplitude and large values of log likelihood ratios (LLRs) at each bit position due to strong neighboring interference. Therefore, a powerful error correcting method will necessarily be demanded for correcting these types of mixed errors at high recording densities.

Low density parity check (LDPC) codes and Reed-Solomon (RS) codes are widely used for correcting these errors. However, it has been shown that RS codes are very robust against hard errors and become weak over AWGN channels. Similarly, LDPC codes perform well in correcting random noises and poorly to hard errors. To solve the main drawback of RS hard decoders, many methods have been proposed to overcome the problems [1], [3]. For small length and high-rate RS codes, perhaps the most impressive results are achieved by the permutation decoding algorithm proposed in [3]. It is shown that the algorithm produces very good error performance and comes extremely close to the maximum likelihood decoder (MLD) within 0.3dB at a bit error rate (bER) of approximately $10^{-5}$. For long code lengths, LDPC codes are commonly used instead of RS codes; however, they are poor at dealing with hard errors caused by media defects, thermal asperity, etc.

The aim of this study is to propose good codes that can perform with the strong ability to correct the three types of errors above. We propose product codes based on binary LDPC codes and binary images of small RS codes. Although, product codes can improve the minimum distance at the expense of code rate [4], they are infrequently used in practical applications such as hard disk drives (HDDs) since their decoding performances are usually poor. To overcome this problem, we proposed two novel algorithms for product codes we refer to as the *error detection algorithm* (EDA) and the *product decoding algorithm* (PDA). We evaluate the performance of the proposed algorithms by applying it to various noise channels. These proposed product codes have both hard and soft iterative decisions.

## II. BACKGROUND

### A. Binary Images of Double-Parity RS Codes

In this subsection, we briefly review the structure of binary images of a double-parity RS code. For complete discussions on the binary images of RS codes, we refer the reader to [1], [3], [4].

Let $\alpha$ be a fixed primitive element in the Galois field $GF(2^m)$. Let $\gamma = [\gamma_1, \gamma_2, \cdots, \gamma_m] = [1, \alpha, \cdots, \alpha^{m-1}]$ be a basis of $GF(2^m)$ over $GF(2)$. Let the code length be $n = 2^m - 1$. The binary image of a $(n, n-2, 3)$ RS code is obtained by representing every codeword $\mathbf{c} = [c_0, c_1, \cdots, c_{n-1}]$ as an $m \times n$ binary matrix.

$$B_M(\mathbf{c}) := \begin{bmatrix} c_{1,0} & c_{1,1} & c_{1,2} & \cdots & c_{1,n-1} \\ c_{2,0} & c_{2,1} & c_{2,2} & \cdots & c_{2,n-1} \\ \vdots & \vdots & \vdots & \vdots & \vdots \\ c_{m,0} & c_{m,1} & c_{m,2} & \cdots & c_{m,n-1} \end{bmatrix},$$

where $c_j = c_{1,j}\gamma_1 + c_{2,j}\gamma_2 + \cdots + c_{m,j}\gamma_m$ for all $j \in \mathbf{Z}_n$. The parity check matrix of the double-parity RS $(n, n-2, 3)$ binary image is represented by the following $2m \times m$ polynomial matrix in the ring $\mathbf{F}_2[x] / (x^n - 1)$.

$$
\begin{bmatrix}
\theta_1(x) & 0 & \cdots & 0 \\
0 & \theta_1(x) & \cdots & 0 \\
\vdots & \vdots & \vdots & \vdots \\
0 & 0 & \cdots & \theta_1(x) \\
\theta_\varepsilon(x)x^{u_1} & \theta_\varepsilon(x)x^{u_2} & \cdots & \theta_\varepsilon(x)x^{u_m} \\
\vdots & \vdots & \vdots & \vdots \\
\theta_\varepsilon(x)x^{u_1+m-1} & \theta_\varepsilon(x)x^{u_2+m-1} & \cdots & \theta_\varepsilon(x)x^{u_m+m-1}
\end{bmatrix} \quad (1)
$$

where $\varepsilon = \alpha^{-1}$, $\theta_\varepsilon(x)$ is known as the **idempotent** [4].

In particular, $\theta_1(x)$ equals $1 + x + x^2 + \cdots + x^{n-1}$ and vector $\mathbf{u} \in \mathbf{F}_n^m$ is computed in Table. 1. [3]

Table. 1. $\mathbf{u}$ vectors computed for $\gamma = [1, \alpha, \cdots, \alpha^{m-1}]$.

|  | Vector $\mathbf{u}$ | Primitive element $\alpha$ |
|---|---|---|
| $GF(2^3)$ | $[2,1,0]^T$ | $\alpha^3 = \alpha + 1$ |
| $GF(2^4)$ | $[2,1,0,14]^T$ | $\alpha^4 = \alpha + 1$ |
| $GF(2^5)$ | $[30,29,28,27,26]^T$ | $\alpha^5 = \alpha^2 + 1$ |
| $GF(2^6)$ | $[4,3,2,1,0,62]^T$ | $\alpha^6 = \alpha + 1$ |

### B. Permutation Decoding Algorithm

In this subsection, an introduction to permutation decoding algorithm is included in order to make this paper self-contained. Further discussion on this topic can be found in [1], [3].

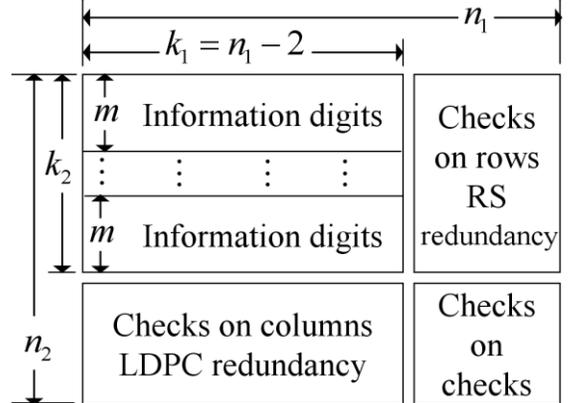

Fig.1. Example of permutation decoding. Each column represents a $GF(2^3)$ symbol in binary notation. Erroneous symbols occur at (1,6), (2,0), and (3,1).

Consider an RS code $(7,5,3)$ over $GF(2^3)$. Let the basis $\gamma = [1, \alpha, \alpha^2]$ where primitive element $\alpha$ satisfies $\alpha^3 = \alpha + 1$. Let the RS codeword $\mathbf{c} = \left(0,1,0,\alpha^5,0,\alpha^2,\alpha\right)$ be transmitted in its binary image as shown in Fig.1. Assume that error positions only occur at (1,6), (2,0), and (3,1). Next, permute the three rows of $B_M(\mathbf{c})$ by (0,4,6)(2,5,3), and (1,4,2)(3,5,6), and (0,4,6)(2,5,3), respectively. It is observed that all the erroneous bits are permuted into the first symbol. Furthermore, the binary image $B_M(\mathbf{c})$ is permuted to $B_M(\mathbf{c}')$, which is also an RS codeword. Therefore, all errors can be corrected using a conventional hard decision decoder. The details of the permutation decoding algorithm are restated below.

**Permutation decoding algorithm [3]**

**Input**: Observations of channel output $\mathbf{y}$. Parameter $\eta$.

**Matrix $\mathbf{S}$ and vector $\mathbf{u}$, basis $\gamma$.**

**Output**: Most-likely codeword in list $L$
1. Perform hard decision decoding on $\mathbf{y}$.
2. **if** $\mathbf{y}$ can be decoded to some codeword $\mathbf{c}$ then store $\mathbf{c}$ in $L$.
3. **forall** $j \in J(\mathbf{y}, \eta)$
4.     Compute a permutation $\pi$ with input $\mathbf{S}, \mathbf{j} - \mathbf{u}$, and $-\mathbf{u}$
5.     Construct $y^{(g)}$ by setting $y_{i,G[i,j]}^{(g)} := y_{i,G[i,\pi(j)]}$.
6.     Compute $(i_0, \tau_0) = \arg\min\left|y_{i,j}^{(g)}\right|$.
7.     Erase 0-th and $\tau_0$-th symbols, decode $y^{(g)}$ to obtain codeword $\mathbf{c}^{(g)}$.
8.     Permute $\mathbf{c}^{(g)}$ with $g^{-1}$ and store in $L$.
9. **end**  □

### III. EFFECTIVE DECODER OF PRODUCT CODES

#### A. Structure of product codes

Fig.2. A product code of binary image of $(n_1, n_1 - 2, 3)$ RS codes in $GF(2^m)$ and $(n_2, k_2)$ binary LDPC codes.

To produce powerful codes, a product of the binary image of $(n_1, n_1 - 2)$ RS code $C_1$ and $(n_2, k_2)$ binary LDPC code $C_2$ is formed. The product code $C_1 \times C_2$ is encoded in two steps. At the first step, each row of the information array is encoded into a RS codeword in $C_1$. At the second step, each of the $n_1$ columns of the array formed in the first encoding step is encoded into an LDPC codeword in $C_2$. This results in a code array of $n_2$ rows and $n_1$ columns, as shown in Fig.2. It should be noted that the information digits are decomposed into many $m$-bit rows corresponding to binary images of RS symbols in $GF(2^m)$ over $GF(2)$. For the long sector format, we investigate a product code of a 2D (1057, 813, 34) projective geometry (PG) LDPC code and a (31, 29) RS code over $GF(2^5)$. The final code-rate is 0.72. Any high code-rate can be constructed using adequate codes.

#### B. Error detection algorithm

We propose an error detection algorithm based on checksum on rows and columns of product codes to enhance the error correction ability of product codes. The details of the *error detection algorithm* (EDA) are described in the following paragraphs.

## Error detection algorithm (EDA)

**Input**: A received product codeword $\mathbf{c}_P$ of size $(n_2, n_1)$. Parameter $n_1$, $k_1$, $n_2$, $k_2$. A parity check matrix $\mathbf{H}$ of the LDPC code.

**Output**: A list of error positions.

1. Checksum on $k_1$ row of product codeword $\mathbf{c}_P$ and output

   a list $L_{\text{row}}$ of erroneous rows.

2. **forall** $i = 1, n_1$

3.    Checksum on the $i$-th column of $\mathbf{c}_P$ using matrix $\mathbf{H}$ and

   output a list of positions $L_{\text{column}}$ that cause errors on all

   checksums.

4.    Output the error positions $(L_i, i)$ where $L_i$ are intersections

   of $L_{\text{row}}$ and $L_{\text{column}}$.

5. **end**          □

It should be noted that the proposed algorithm can be regarded as a generalization of the error detection method in single parity check (SPC) product codes using the parity parts of RS codes and LDPC codes. Moreover, we empirically found that the detection ability of our algorithm strongly depends on the minimum Hamming distance of LDPC codes, such as in the case using (1057, 813) PG-LDPC codes with a minimum distance $d_{\min} = 34$. Our method works perfectly even in various noise channels.

*Example 1*: Consider an example product code of a binary image of RS (7, 5, 3) and a Hamming code (7, 4, 3) with the parity matrix below.

$$\mathbf{H} := \begin{bmatrix} h_1 \\ h_2 \\ h_3 \end{bmatrix} = \begin{pmatrix} 1 & 1 & 0 & 1 & 1 & 0 & 0 \\ 1 & 0 & 1 & 1 & 0 & 1 & 0 \\ 0 & 1 & 1 & 1 & 0 & 0 & 1 \end{pmatrix}.$$

We re-use the RS codeword as shown in Fig.1. After two-step encoding, we obtain a product codeword $\mathbf{c}_{7\times7}$, whose first three rows contain information digits from the RS codeword. It should be reminded that row 4 of $\mathbf{c}_{7\times7}$ contains redundant information digits only for this example. Suppose $\mathbf{c}_{7\times7}$ has errors only at (1,6), (2,0), and (3,1). In order to detect errors, we first check on the first three rows of $\mathbf{c}_{7\times7}$ and obtain $L_{\text{row}} = \{1, 2, 3\}$. We then check on the first column of $\mathbf{c}_{7\times7}$. From Table.2, it can be seen that parity-check failures occur on check row $h_1$ and $h_3$; hence, the possible error locations could be $L_{\text{column}} = \{2, 4\}$, which are the common "1" in both $h_1$ and $h_3$. Therefore, it can be concluded that the error location occurs at (2,0). The others (3,1) and (1,6) can be detected similarly. □

Although, this example was applied to a small code length, our method can be used for all $(n_1, n_1 - 2, 3)$ RS codes and the entire parity check matrix $\mathbf{H}$. It should be noticed that our method does not requires $GF(2^m)$ operations, hence it is effective and low in complexity.

Table 2. Product codeword $\mathbf{c}_{7\times7}$ with error positions at (1,6), (2,0) and (3,1).

| position | 0 | 1 | 2 | 3 | 4 | 5 | 6 | sum on rows |
|---|---|---|---|---|---|---|---|---|
| rows of | 0 | 1 | 0 | 1 | 0 | 0 | **1** | = 1 → error |
| (7, 5, 3) | **1** | 0 | 0 | 1 | 0 | 0 | 1 | = 1 → error |
| RS code | 0 | **1** | 0 | 1 | 0 | 1 | 0 | = 1 → error |
| redundant | 1 | 1 | 1 | 1 | 1 | 1 | 1 | no check sum |
| parity of | 1 | 0 | 1 | 1 | 1 | 1 | 0 | on row 4 to |
| Hamming | 1 | 0 | 1 | 1 | 1 | 0 | 1 | row 7. |
| code (7,4) | 1 | 1 | 1 | 1 | 1 | 0 | 0 | |
| check | 1 | 0 | 0 | 0 | 0 | 0 | 1 | → detect |
| on | 0 | 1 | 0 | 0 | 0 | 0 | 1 | errors at (1,6), |
| columns | 1 | 1 | 0 | 0 | 0 | 0 | 0 | (2,0), (3,1) |

### C. Decoding algorithm for product codes

Permutation decoders show the ability to correct more hard errors than conventional hard decision decoders [3]. However, they require adequate erasure information of bit locations and the length of RS codes to be quite small. For practical applications, our proposed structure resolves these problems. Based on our EDA method in the previous subsection, a novel decoding algorithm for product codes is proposed and denoted as the *product decoding algorithm* (PDA). Fig.3. describes the flow diagram of the product decoding algorithm.

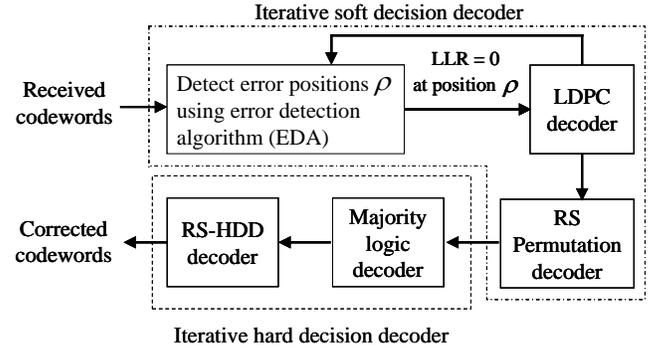

Fig.3. Product decoding algorithm for product codes from binary image of $(n, n - 2, 3)$ RS codes and binary LDPC codes.

## Product decoding algorithm (PDA)

**Input**: Received codewords of a product code based on binary image of RS $(n, n - 2, 3)$ code and a binary LDPC code.

**Output**: Corrected codewords.

1. Detect error positions $\rho$ using proposed error detection algorithm.

2. Set log likelihood ratio (LLR) at position $\rho$ equal to zero.

3. Iterative decoding on columns of the product code by the sum product algorithm (SPA); repeat step 1.

4. After several iterations, apply RS permutation decoding on information rows of the product code.

5. Decode on columns of the product code using the majority logic decoder.

6. If some errors exist, apply hard decision decoder on rows of the product code and output the corrected codewords. □

It can be clearly seen that our EDA is necessarily used to detect and erase hard errors before executing SPA decoding on columns of product codes. This is because LDPC decoders perform poorly at correct SHEs. Simulation results show that the total numbers of errors ware reduced quickly by the SPA algorithm after applying the EDA method. For decoding on rows of product codes, RS permutation decoders are used to eliminate the remaining erasure errors. To correct two errors from occurring in the same row of the product code, we use an iterative hard decision decoder including LDPC majority logic decoding on columns and RS-hard decision decoding on rows of the product codes as shown in Fig.3.

## IV. PERFORMANCE EVALUATION

### A. Product codes of LDPC codes and RS codes

In our evaluation, a double-parity (31,29) RS code over $GF(2^5)$ is used as the inner codes. For the outer codes, we investigate a 2-dimensional (1057, 813, 34) projective geometry (PG) LDPC code given in [2], with a column weight of 33 and a minimum distance of 34. The length of the final product code $C_{PG}$ is 32767 and the code-rate is 0.72. The maximum number of SPA iterations is set to 20 for all LDPC decoders.

### B. Performance Evaluation of Product Codes

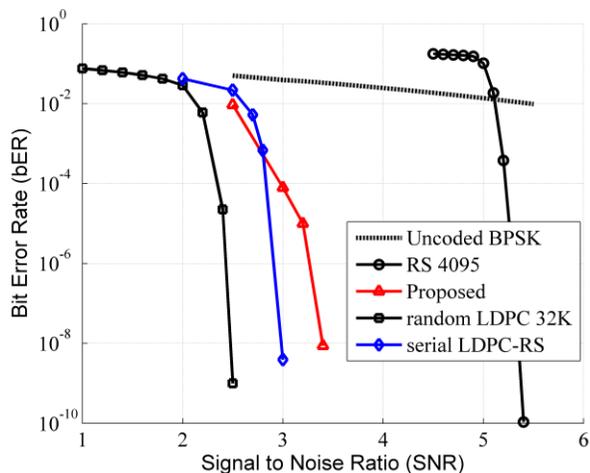

Fig.4. bER of the product code $C_{PG}$ based on a (1057, 813) PG-LDPC code and a (31,29) RS code over AWGN channels.

Fig.4. compares the bER performance of the $C_{PG}$ code with an equivalent Reed-Solomon code, random LDPC code and serial concatenation LDPC-RS code over AWGN channels. The RS code is a (4095, 2947) over $GF(2^{12})$ that can correct up to 574 symbol errors. The LDPC code is a (32767, 23592) random binary LDPC code with no 4-cycle. The equivalent serial concatenation of LDPC and RS code has the same code length and 20% and 8% of code length are used as the parity parts of LDPC outer and RS inner code, respectively. The $C_{PG}$ code is decoded by the PDA method. Fig.4. shows that the $C_{PG}$ code outperforms the same code-rate RS code decoded using the hard decision decoding algorithm by 1.9dB at a bER of approximately $10^{-8}$. Although its bER performance is slightly degraded compared to that of the LDPC code and the serial concatenation LDPC-RS code over AWGN channels,

our product code can be used in real applications of ultra-high density HDDs where error performance is dominated by SHEs and long burst errors. Fig.5 presents the bER performance in the case of mixed error channels including random noises and SHEs where SNR is set at 5dB and 150 to 400 SHEs are randomly generated with amplitudes ranging from 1.0 to 1.5. When the amplitude of SHEs equals 1.5 as shown in Fig.5, the bER performance of $C_{PG}$ shows a significant improvement compared to the equivalent random LDPC code and the serial concatenation LDPC-RS code. Moreover, our code can correct any two arbitrary columns containing long burst errors of length 1057 based on the ability of the RS code. Therefore, the proposed code will be useful for the implementation of future HDDs.

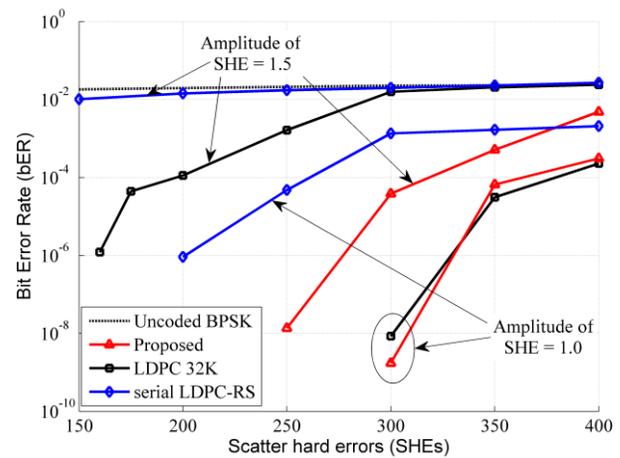

Fig.5. bER of the product code $C_{PG}$ based on a (1057, 813) PG-LDPC code and a (31,29) RS code at SNR = 5dB and hard errors from 150 to 400.

## V. CONCLUSIONS

The proposed algorithms using parity check on rows and columns of product codes were able to improve the bER performance in the presence of SHEs, AWGN errors and long burst errors of length up to 1057 based on the ability of the RS code. Therefore, the proposed code might become a key step to practical implementation of future HDDs.

## ACKNOWLEDGMENT

This study was supported by the Japan Society for the Promotion of Science (JSPS) for Scientific Research 21560418, the Storage Research Consortium (SRC), and the NEDO project.